\documentclass[aps,prb,groupedaddress]{revtex4}

\usepackage{graphicx}
\begin{document}


\title{Conceptual and Epistemological discussions on Quantum Mechanics in a Virtual Laboratory}

\author{F. Ostermann}
\author{S. D. Prado}
\affiliation{Instituto de F\'{\i}sica, Universidade Federal do Rio
Grande do Sul, 91501-970 Porto Alegre, RS, Brazil}

\date{\today}

\begin{abstract}
We argue here for a more conceptual or qualitative approach in the
introductory teaching of Quantum Physics which is built on the
basis of epistemological and ontological discussions and as such
is a valuable tool mainly in the initial and continued formation
of school teachers. We illustrate our point with the analysis of
undulatory and corpuscular phenomena of a single photon in a
virtual Mach-Zehnder Interferometer - an experimental setup
similar to double slits device but simpler - using key ideas  of
the most known interpretations of quantum formalism, including the
many worlds interpretation.
\end{abstract}


\maketitle

\section{Introduction}

There are several factors prompting new strategies in the teaching
of physics. First, the upsurge in interest in questions concerning
fundamental aspects of Quantum Physics which has been stimulated
by the experimental progress of the last two decades. It is now
becoming possible to apply Quantum Mechanics(QM) particularly in
the fields of Computater Science and
Cryptography\cite{Zeilinger_1996}. There is then a need for
introducing the teaching of Quantum Mechanics for people with
little or no background in Physics. For this technological
motivation some authors\cite{Mermin_2003,Grau_2004} suggest
teaching students just enough operational quantum mechanics to
understand and develop algorithms in quantum computation and
quantum information theory. Second, there are initiatives of
introducing quantum mechanics in the upper secondary
school\cite{Olsen_2002, Michelini_2000} based on the fact students
should have some understanding of how fundamentally this part of
physics differs from classical physics. Finally, some authors
point towards a greater focus on conceptual understanding and the
cognitive skills\cite{Kalkanis_2003} required to understand and
apply physics concepts and the use of technology as a teaching
tool to revert the decline in the number of students choosing
physics as a major field of study\cite{Thacker_2003}.

Our claim here is that special attention has to be given to
problems of conceptual understanding and interpretation of Quantum
Physics since students from other careers or Physics teachers have
not enough time in their standard courses to devote a great deal
of effort to trying to understand the formal structure of quantum
theory. The long training in the quantum formalism is a requisite
for any theoretical physicist\cite{Deutsch_book_51}, but the
formalism can not be alone a tool in teaching introductory quantum
mechanics in more general situations. Also, a conceptual approach
as an introduction to the formalism should be used as a motivation
even in regular physics courses. By  formalism or formal structure
we mean the theory's mathematical structures that is well
supported by experimental evidence, while an interpretation of
quantum mechanics is an attempt to answer what exactly  quantum
mechanics is talking about.

For our purposes we have to be more precise about the kind of
picture an interpretation provides. Like other theories, Quantum
Physics can be formalized in terms of several axiomatic
formulations. Accordingly to Jammer\cite{Jammer_book}, we can
distinguish at least two components in a physical theory
\texttt{T}: (1) an abstract formalism \texttt{F} and (2) a set of
correspondence rules \texttt{R}. The formalism \texttt{F}, the
logical backbone of the theory, is an axiomatized
 deductive calculus in general devoid of any empirical sense.
Although the formalism might contain words like {\em particles}
and {\em state} that may suggest physical reality, these terms do
not have any other meaning besides the place they occupy in the
context of \texttt{F}.  For \texttt{F} to be physically
meaningful, some formula need to be correlated with observable
phenomena and empirical operations. These correlations are
expressed through correspondence rules \texttt{R}. \texttt{F}
without \texttt{R} is a game without physical context. We denote
\texttt{F}$_R$ the formalism \texttt{F} when entwined by the
correspondence relations \texttt{R}. It is the interpretation of
\texttt{F}$_R$ which gives rise to the philosophical problems like
the ontological problem of physical reality. As an example of
\texttt{F}$_R$ in  QM, there is the statistical interpretation
introduced by Max Born in 1926. This interpretation relates the
wavefunction modulus squared
 to probability densities of finding an electron, for
example,  in a particular region of space. It relates the abstract
elements of the theory, such as the wavefunction, to operationally
definable values, such as probabilities. This interpretation is a
consensus among physicists, since predictions obtained via this
rule are in agreement with experiments to an excellent degree of
accuracy\cite{Zeilinger_1999}. Actually, it is not the teaching of
this kind of interpretation which is embodied in the theory we
will explore in this paper.

There is another class of interpretation in QM which is not an
\texttt{F}$_R$. These are lines of philosophical thoughts or
school of thoughts that coexist. Each one has elements that escape
from a complete and detailed description of an experiment and that
are not, by any means, verifiable in laboratory, at least
immediately, since they all predict the same result for a given
experiment\cite{Zeilinger_1996}. These interpretations deal with
the ontology of QM, that is, with the nature attributed to a
quantum object: corpuscular, undulatory or dualist (wave and
particle) and with epistemological attitudes: realist (there is a
world independent of an observer who perceives it) or positivist
(all our knowledge derives from our senses). The difficulties
about interpretations in QM are twofold, to say the least: first,
it is about the way the theory is related to the physical
phenomena and second, it is still missing an appropriate ontology.

 The Copenhagen interpretation still appears to be the most
popular one among scientists\cite{Tegmark_1998}, but it is also
true that most physicists consider non-instrumental questions (in
particular ontological questions) to be irrelevant to physics or
even that an interpretation is nothing more than a formal
equivalence between sets of rules for operating on experimental
data, thus  suggesting that the whole exercise of interpretation
is unnecessary\cite{Fuchs_2000}. As evidenced by
Jammer\cite{Jammer_book}, the success of QM has lead the majority
of physicists to be more interested in practical problems or
applications in such a way that it is not an exaggeration to say
that most of the textbooks deal almost exclusively with one or two
different formalisms used to solve specific problems, leaving no
room  to epistemological questioning. They fall back to the famous
view of Paul Dirac: ``Shut up and calculate" attributed to Richard
Feynman\cite{Mermin_2004}.

If teaching were  to be based only on the formal aspects of QM, we
would never have  to face the problems with interpretation. These
difficulties seem to appear justly where concepts like
understanding and meaning are required\cite{Bastos_2003}. A more
conceptual and qualitative approach of QM is a really valuable
tool for the initial and continued formation of school teachers.
Also, possible didactic transpositions to the introduction of QM
on a pre-university level depend on a solid conceptual background
that can only be built on the basis of epistemological and
ontological discussions. The conceptual and the epistemological
are intertwined when it is asked what QM could represent for our
worldview. The conception of  quantum objects - which can be
illustrated by a simple question about the nature of a photon -
can only be made in the light of philosophical posture which, if
not explicit, can drive to ingenuous  or uncritical views or to
the idea that only a single interpretation is possible. Moreover,
students' own epistemological views should not be neglected, since
some of the difficulties they face in understanding  QM concepts
are of philosophical nature\cite{Lising_2004}. In conclusion, any
attempt of conceptual discussion on QM brings forth elements of
its epistemology, which if not present, makes  the understanding
devoid of any meaning.

This paper is organized as follows: in section II one finds
introductory notes of four lines of interpretations of QM. It is
not our primary objective in this paper to discuss the
interpretations of QM in full detail, but to illustrate their use
as a way of enriching the teaching of QM;  in section III
interpretations of interference phenomena  in a virtual
Mach-Zehnder Interfermometer available on the
worldweb\cite{Muller_2002} are discussed and finally, in section
IV one finds some concluding remarks.

\section{Interpretations of QM}

One of the main difficulties one finds in teaching introductory QM
is to bring together, in terms of understanding, antagonist ideas
like the concept of wave - a non-localized phenomena - and the
concept of particle - an entity. Although the undulatory theory of
light describes the interference pattern in the famous Young
experiment with coherent superposition, Einstein's  explanation
for the Photoeletric Effect requires  light to be composed of
indivisible corpuscles of light called photons. Moreover,
analogous double slit experiment for electrons, neutrons and more
recently molecules like the C60-Fullerene\cite{Arndt_1999} also
exhibit interference fringes. Then, to explain these results it is
necessary to recall  de Broglie's  Postulate that states that
microscopic particles can also behave like waves sometimes. This
double character is widely known as wave-particle duality and it
is mostly associated with light\cite{Eisberg_book}.

There are no classical counterparts to this phenomenon and the
result of a detection or measurement  can present different
epistemological views although they are internally consistent. To
give a highlight to these differences in  a virtual experiment
(section \ref{IntMZI}), we employ a few key points of four
interpretations: i) Undulatory Interpretation or Schr\"odinger
wavefunction, ii) Complementarity Principle  or School of
Copenhagen, iii) Hidden-variables or Dualist-realist
Interpretation and iv) Many Worlds Interpretation. To partly
justify our choices, we have taken into consideration the fact
that the Many-worlds interpretation is currently in the media and
 that throughout much of the twentieth century Copenhagen Interpretation has
had obvious majority acceptance among
physicists\cite{Tegmark_1998,Mermin_2004}.

i) Undulatory Interpretation  (a realist interpretation proposed
by E. Schr\"odinger in 1926) - When Schr\"odinger postulated the
equation known as Schr\"odinger Equation and its boundary
conditions in 1926, he established a formalism in terms of
wavefunctions (or states)  $\Psi$ that have identical or similar
situations in Classical Mechanics\cite{Jammer_book}. For didactic
reasons this formalism has been widely adopted in textbooks since
then. In this interpretation,  physical reality is attributed to
the state independently of any measurement and without any
additional hypothesis that there is anything else besides the
quantum formalism. However, the state is not a directly accessible
reality, but under Born's guidance, it establishes probabilities
that evolve over time just like wavefunctions. $\Psi$ is the
central object in this interpretation that presents no
difficulties in explaining undulatory phenomena. In short, the
photon is taken as a wavepacket.

ii) Complementarity Principle or School of Copenhagen (a
dualist-positivist interpretation  formulated  by N. Bohr and W.
Heisenberg in 1927) - Despite an extensive literature which refers
to the Copenhagen interpretation\cite{Bunge_2003}, the original
formulation
 has led to several variants, making it difficult to establish how
exactly this interpretation is stated. For our purposes, we can
say that the Principle of Complementarity formulated by Bohr in
1927
 establishes that it is not possible, in a single experiment,
 a corpuscular  and an undulatory description
simultaneously. This interpretation is considered
dualist-positivist since it admits the wave-particle duality, but
it also emphasizes that the theory can only explain the results of
experiments that should be predicted, and therefore additional
questions are not scientific but rather philosophical. In contrast
to Schr\"odinger undulatory interpretation, the state here is a
mere mathematical instrument that permits predictions of
measurement, but it is not provided with any physical reality.

iii) Hidden-variables  (a dualist-realist interpretation proposed
by L. de Broglie in 1925 and reformulated by D. Bohm in 1952) -
This is an interpretation of realistic content in which a quantal
system is described not only by its state $\Psi$, but also with
the help of additional hidden variables labelled by a parameter
$\Lambda$ that contain information on the particle: energy,
position and velocity. The state $\Psi$ is a guide-wave or a field
of quasi-probabilities that drives the particle\cite{Jammer_book}.
$\Psi$ and $\Lambda$ together establish where a photon will be
detected, for instance. Briefly, the photon is a particle  to
which is associated a guide-wave.

iv) Many Worlds Interpretation (a realist interpretation proposed
by H. Everett III in 1957) - This is essentially an interpretation
classified as undulatory-realist. Its origin is historically
related to the development of Relativistic Quantum Theory  where
the idea of a wavefunction for the whole universe (an isolated
system) has posed some difficulties to the dualist prevailing
interpretations. It was Hugh Everett III who proposed an idea that
in a measurement all possible outcomes of the observable are
obtained simultaneously, but in parallel worlds - these worlds are
complex subsystems causally connected that can be forced to
interfere  with one another. In this interpretation one does not
recur to the postulate of wavefunction collapses, differently from
all the undulatory interpretations we have previously mentioned.
As a realist undulatory interpretation, the photon is again
considered a wavepacket.

\section{\label{IntMZI}Interpretations of a Virtual Experiment}

\begin{figure}
\centering
\includegraphics[width=8.cm,height=6.0cm]{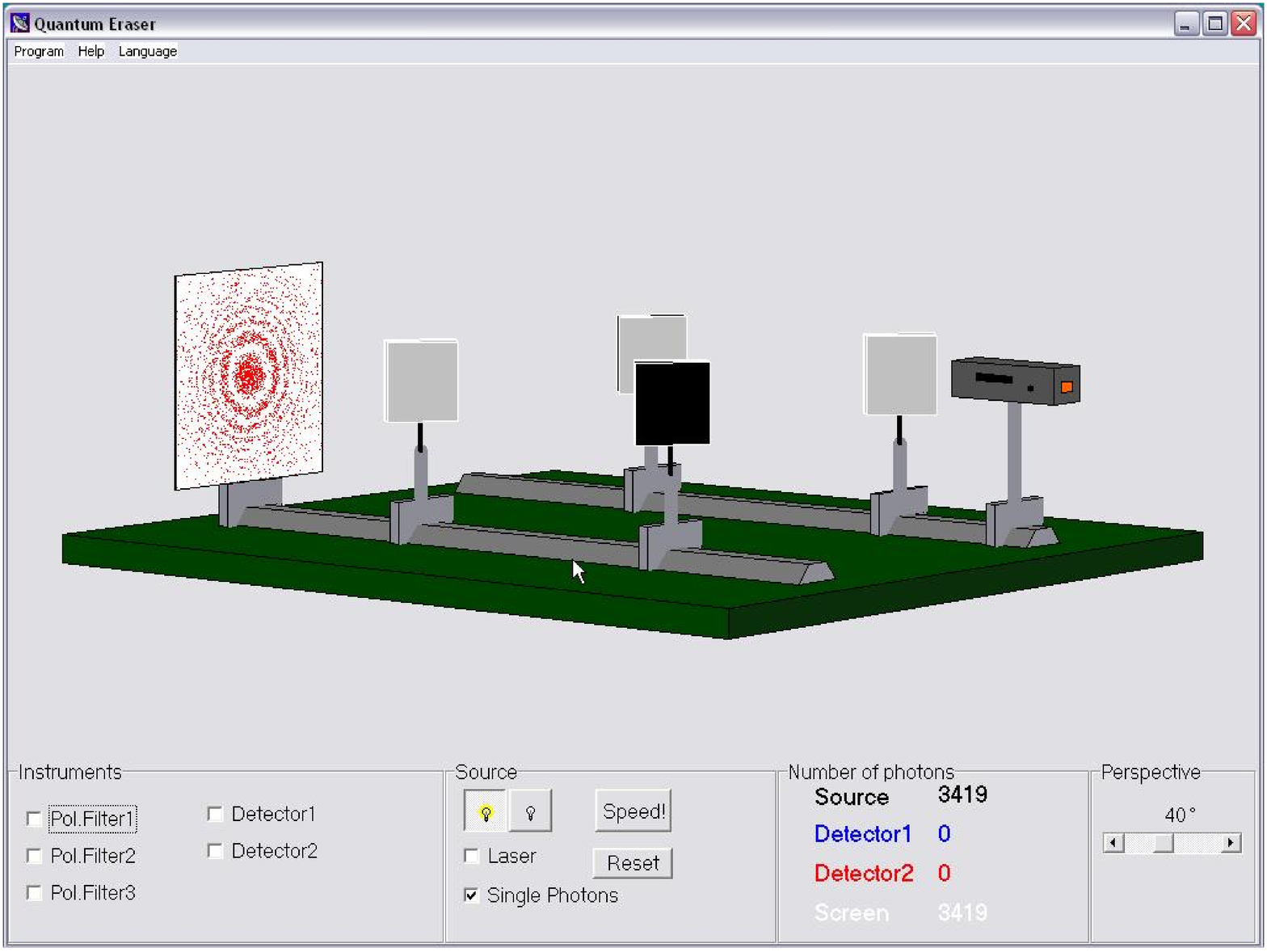}
\hspace{0.1cm}
\includegraphics[width=8.0cm,height=6.0cm]{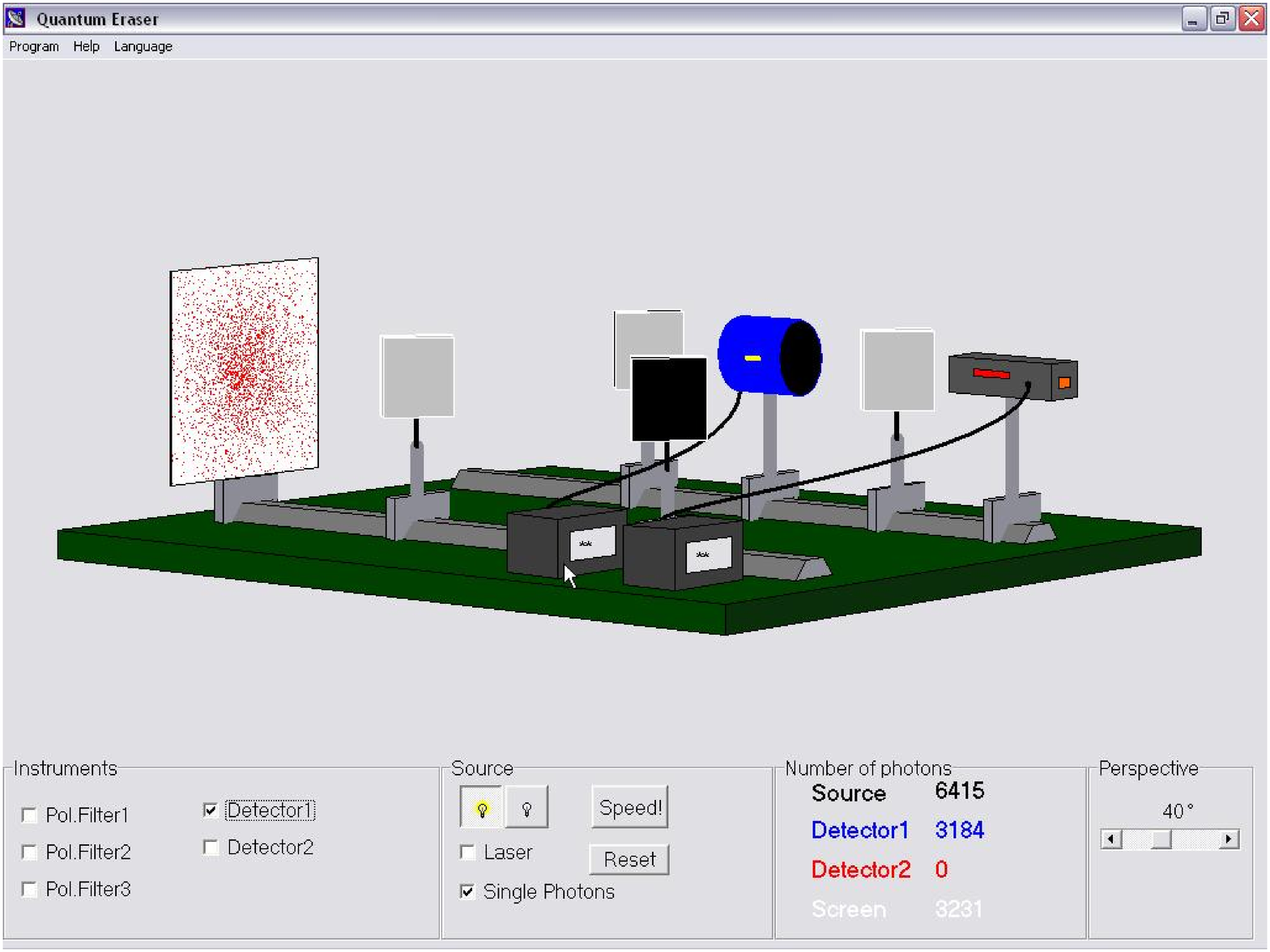}
\caption{\label{figure1} Virtual Mach-Zhender Interferometer using
a single photon source (free code\cite{Muller_2002}). On the left,
an undulatory pattern is clearly seen after 3,419 photons were
emitted. On the right, with an additional detector in the upper
arm  6,415 photons built up a corpuscular pattern. 3,184 photons
were detected  and 3231 hit the screen.}
\end{figure}

The Mach-Zhender Interferometer (MZI) sketched in figure
(\ref{figure1}) is an experimental device totally analogous to the
double slit experiment where it is possible to observe wave
interference. It has been mentioned quite often in the
introductory QM literature, given its usefulness  as a pedagogical
tool\cite{Muller_2002,Pessoa_1997,Pessoa_book}. It is composed of
two half-silvered mirrors or beam splitters that transmit 50\% of
incident light (upper arm) and reflect the other half (lower arm),
plus two usual mirrors and a screen. When a laser source is active
a pattern of rings can be seen. It is relevant to recall two
different regimes here: the classical regime described by the
undulatory theory of light that one would obtain in any
conventional teaching laboratory with a He-Ne
laser\cite{Eisberg_book} and the quantal regime, in which the beam
intensity is diminished at the level of emission of a single
photon at a time - a monophotonic regime. This is the regime our
discussions are centered on. It is worth to say  this
simulator\cite{Muller_2002} offers a much wider number of
experiments with light and photons that can be discussed in a
classroom. We have selected two of the most simple cases to show
how interpretations allied to virtual simulations can be a
powerful tool in teaching effectiveness.

\noindent {\bf Undulatory pattern}

\noindent {\em As the number of photons that have left the single
photon source one at a time increases, a pattern of rings is
gradually built on the screen} (figure (\ref{figure1}) on the
left).

The interesting results from the QM view emerge in the limit of
one photon at a time. In this case, detection is punctual and
restricted to certain places over the screen. QM formalism
establishes that the half-silvered mirror nearest to the source
places the photon in a superposition of states - the state
describing a photon running along the upper arm  and the state for
running along the lower arm. As the chances of being reflected or
transmitted are the same, the two states are equally probable in
terms of Born Postulate.

In the classic experiment (laser source), when light passes
through MZI  onto a screen, alternate bands of bright and dark
regions are produced. These can be explained as areas in which the
light waves reinforce or cancel. With a single photon source only
one photon enters the interferometer each time. In performing the
experiment, a photon hits the screen one at a time. However, when
one totals up where the photons have hit, one will see
interference patterns that appear to be the result of interfering
waves even though the experiment dealt with one particle at a
time.

i)Undulatory Interpretation - The photon is a wavepacket that is
divided in  transmitted and reflected parts in the first
half-silvered mirror (mirror nearest to the single photon source).
These two waves travel along  upper and lower arms accumulating
phases to recombine again in the second half-silvered mirror in a
constructive or a destructive superposition that determines where
the photon hits the screen. The detection over the screen is still
punctual, as if the photon were a corpuscle, but this detection
does not represent any difficult for this interpretation, since a
wavefunction collapse  is assumed in the process of measurement
(interaction with the screen). So, the detection on the screen
happens only in regions where waves satisfy the condition of
constructive interference. Within this interpretation questioning
the way a photon is travelling is nonsense.

ii) Complementarity Principle - The result of the virtual
experiment shows a pattern of interference so the photon  behaved
as a wave. It does not even make sense to ask what a photon is
before it hits the screen. Physics is the science of outcomes of
measurement processes and speculation beyond that is not
justified. The act of measuring causes an ``instantaneous''
collapse of the wavefunction. This means that the measurement
process randomly picks out exactly one of the two possibilities
allowed for by the wavefunction.

iii) Hidden-variables - In this interpretation there is not any
difficult about the fact that detection on the screen is punctual.
The undulatory behavior and corpuscular characteristics are all
together in the description of the experiment. In the MZI, the
wavepacket  is partially transmitted and partially reflected in
the first half-silvered mirror. Just like in the undulatory
interpretation, these waves will recombine again after they reach
the second half-silvered mirror in a way that constructive or
destructive interference will determine the probabilities that the
photon will be detected on a particular region of the screen.
However, the particle photon will follow one of the two arms
according to hidden variables - this information is not accessible
in the experiment. The detection will happens at those places
where waves interfere constructively.

iv) Many worlds Interpretation - When a photon reaches the first
half-silvered mirror the world splits into two identical new ones
- one  with the photon travelling in the upper arm while in the
other, the photon goes along the lower arm. Observer in one world
ignores the simultaneous existence of the other which is
unaccessible unless there are interference of  one with another.
For that to happen, all atoms, subatomic particles, photons and
other degrees of freedom have to be in the same state, meaning
that they have to be in the same place or to have a significant
superposition. This superposition condition is made through the
second half-silvered mirror.  So, after a passage through the
second mirror, a photon will be detected only in regions of
constructive interference on the screen.

\noindent {\bf Corpuscular behavior}

\noindent {\em  With a detector at the upper arm, there is a click
in the detector or a photon hitting the screen. After many photons
have gone through the interferometer the pattern built up on the
screen is corpuscular} (figure \ref{figure1} on the right).

i)Undulatory Interpretation -  The photon that enters the MZI
splits into a transmitted wave that follows  the upper arm and a
reflected wave that follows the lower arm, so that its behavior is
described by a non-localized wave function which is travelling
through both arms at the same time. However, a measurement forces
the wavepacket to collapse and the detection happens in the
detector where a collapse of the wavefunction is likely.

ii) Complementarity Principle - In the virtual experiment, a
photon is detected in the counter or it hits the screen.  Then, as
the result of the experiment shows, the photon behaved like a
particle that had been transmitted or reflected in the first
mirror with equal probabilities.

iii) Hidden-variables - In this interpretation the wave-guide or
associated wave that travels
 with the photon divides itself into a transmitted and a reflected
 part in the first half-silvered mirror but the photon follows one of
 the two
 trajectories. The wave-guide dictates where the photon will be detected.
 Again, there are some difficulties in explaining a corpuscular pattern without
 taking into account the peculiarities of the measurement process or wavefunction
 non-local collapses.

iv) Many worlds Interpretation - In this interpretation, the world
is split into two when a photon reaches
 the first mirror.   In this way, if the photon is detected in one of the arms
 in one world, it hits the screen in a second world, but this result in
 unaccessible to the former observer, just as the photon hitting the detector is
 a result unaccessible to the second observer. The photon is still
 a wavepacket, but no interference fringes are observed. The
 addition of a detector in one of the arms destroys the perfect
 match of all degrees of freedom that should be in superposition to show
 interference fringes.

\section{Final Remarks}

The strangeness of QM  instigates questions that varies from a
lack of a proper language - words that have yet to be invented -
to describe microscopical scale behavior  to the need of a
non-classical logic - a quantum
logic\cite{Deutsch_book_51,Tegmark_2001}. Exclusively undulatory
interpretations show some difficulties in explaining corpuscular
phenomena just like a exclusively corpuscular interpretation
should face difficulties in  explaining an interference pattern.
The complementarity interpretation combines both characteristics -
undulatory and corpuscular - but in a way that one excludes the
other. Although the many worlds interpretation gets rid of
wavefunction collapses, it implies the existence of parallel
universes that are connected only weakly through interference
phenomena\cite{Deutsch_book_51}.

It has to be stressed that our point here is not to be in favor of
one particular interpretation over another but to advocate that
more conceptual and qualitative approaches built on the plurality
of interpretations are pedagogical tools in the introductory
teaching of QM. Besides the fact that for many students the
formalism without some interpretation remains abstract mathematics
that is very unlikely to remain in the long run, it is also a
relief for many students to find out that their own
epistemological views may be shared by many famous scientist.
Recent polls in classrooms (motivated by \cite{Tegmark_1998}) have
shown that students voluntarily join themselves in groups in
support of one or another interpretation. The concepts they have
assimilated (or not) are evidenced  when groups debate in favor of
their choices. Also, in favour of a more conceptual and
qualitative approach in introductory courses, is the fact that
students meet in class themes like parallel universes that are
often in the media or the worldweb and  are not or are scarcely
presented in textbooks. The discussion of interpretations allied
to a virtual experiment\cite{Muller_2002} of the photon in an MZI
is an important contribution to the literature devoted to initial
and continued formation of Physics teachers.

This is a current research project in  Education of Contemporary
Physics and Epistemological Foundations. It is partially supported
by CNPq. We thank SR Dahmen for English improvements.


\end{document}